\RequirePackage{ifpdf}
\documentclass{PoS}

 \usepackage{graphicx}
\ifpdf
\usepackage{epstopdf}
\fi
\usepackage[multidot]{grffile}

\usepackage{fixltx2e}

 \usepackage{graphics}
 \usepackage{caption}
 \usepackage{epsfig}
 \usepackage{longtable}
 \usepackage{latexsym}
 \usepackage{amsmath}
 \usepackage{amsthm}
 \usepackage{amsfonts}
 \usepackage{amssymb}
 \usepackage{dsfont}
 \usepackage{bm}
 \usepackage{subfigure}
\MakeRobust{\eqref}
\renewcommand{\>}{\rangle}
\renewcommand{\dag}{^\dagger}

\newcommand{\<}{\langle}

\newcommand{\be}{\begin{equation}}
\newcommand{\ee}{\end{equation}}
\newcommand{\bi}{\begin{itemize}}
\newcommand{\ei}{\end{itemize}}
\newcommand{\bey}{\begin{eqnarray}}
\newcommand{\eey}{\end{eqnarray}}

\title{Chiral condensate from the Banks-Casher relation}
\ShortTitle{Chiral condensate from the Banks-Casher relation}

\author{\speaker{Georg P.~Engel}
\\
Dipartimento di Fisica, Universit\`a Milano-Bicocca, \\
and INFN, Sezione di Milano-Bicocca, \\
Piazza della Scienza 3, 20126 Milano, Italy \\
E-mail: \email{georg.engel@mib.infn.it}
}
\author{Leonardo Giusti\\
Dipartimento di Fisica, Universit\`a Milano-Bicocca, \\
and INFN, Sezione di Milano-Bicocca, \\
Piazza della Scienza 3, 20126 Milano, Italy \\
E-mail: \email{leonardo.giusti@mib.infn.it}
}
\author{Stefano Lottini\\
DESY, Zeuthen, \\
Platanenallee 6, 15738 Zeuthen, Germany \\
E-mail: \email{stefano.lottini@desy.de}
}
\author{Rainer Sommer\\
DESY, Zeuthen, \\
Platanenallee 6, 15738 Zeuthen, Germany \\
E-mail: \email{rainer.sommer@desy.de}
}

\abstract{
We report on our ongoing project of determining the chiral condensate of two-flavor QCD from the Banks-Casher relation. 
We compute the mode number of the O(a)-improved Wilson-Dirac operator for several values of $\Lambda$, and we discuss different fitting strategies to extract the chiral condensate from its mass and $\Lambda$ dependence.
Our preliminary results haven been obtained at two different lattice spacings by using CLS-configurations.
}

\FullConference{The XXXI International Symposium on Lattice Field Theory, Lattice 2013\\
		July 29 - August 03, 2013\\
		Johannes Gutenberg Universit\"at Mainz, Germany}

\begin{document}
%%%%%%%%%%%%%%%%%%%%%%%%%%%%%%%%%%%%%%%%%%%%%%%%%%%%%%%%%%%%%%%%%%%%%%%%%%%%%%%

%%%%%%%%%%%%%%%%%%%%%%%%%%%%%%%%%%%%%%%%%%%%%%%%%%%%%%%%%%%%%%%%%%%%%%%%%%%%%%%
\section{Introduction}
\label{sec:intro}
%%%%%%%%%%%%%%%%%%%%%%%%%%%%%%%%%%%%%%%%%%%%%%%%%%%%%%%%%%%%%%%%%%%%%%%%%%%%%%%
\noindent
The chiral condensate provides an order parameter for the dynamical breaking of chiral symmetry in QCD. 
A comprehensive list of recent related lattice QCD results is found in Ref.~\cite{Colangelo:2010et}. 
The present work exploits the Banks-Casher relation  \cite{Banks:1979yr}, which links the chiral condensate $\Sigma$ to the low end of the spectral density $\rho$ of the massless Dirac operator, 
\bey
\Sigma 						&=& \pi \lim_{\lambda \to 0} \lim_{m \to 0} \lim_{V \to \infty} \rho(\lambda,m) \;,\\
\text{with} \qquad \rho(\lambda,m) 	&=& \frac{1}{V}\sum_{k=1}^{\infty} \< \delta(\lambda-\lambda_k) \> \;, 
\eey
$m$ the current quark mass, $\lambda_k$ the eigenvalues of the massless Dirac operator and $V$ the four-volume. 
The spectral density is a renormalizable quantity in QCD and can be computed on the lattice numerically \cite{Giusti:2008vb}.
Still, for numerical evaluation it is more convenient to consider the mode number $\nu(\Lambda,m)$ of the massive hermitian operator 
$D\dag D-m^2$ with eigenvalues $\alpha\leq\sqrt{\Lambda^2+m^2}$, which is renormalization-group invariant, 
\bey
\nu(\Lambda,m)			&=& V \int_{-\Lambda}^{\Lambda} d\lambda \rho(\lambda,m) \\
\nu_R(\Lambda_R,m_R)		&=& \nu(\Lambda,m)\;.
\eey
The procedure was shown to work in Ref.~\cite{Giusti:2008vb} and applied to twisted--mass fermions in Ref.~\cite{Cichy:2013gja}. 
In the following, we drop the subscript $R$, and all quantities are renormalized unless stated otherwise. 
To extract the chiral condensate, we choose to define the effective condensate, 
\be
{\widetilde \Sigma}(\Lambda_1,\Lambda_2,m) 	= \frac{\pi}{2 V} \frac{\nu(\Lambda_2)-\nu(\Lambda_1)}{\Lambda_2 - \Lambda_1} 
\;\; \xrightarrow{V\rightarrow\infty;m,\Lambda_{1,2} \rightarrow 0} \;\; \Sigma
\;,
\ee
which agrees with $\Sigma$ as $\Lambda_{1,2}$ and $m$ go to zero. 
This definition removes any threshold effects as long as $\Lambda_{1,2}$ are chosen large enough.

%%%%%%%%%%%%%%%%%%%%%%%%%%%%%%%%%%%%%%%%%%%%%%%%%%%%%%%%%%%%%%%%%%%%%%%%%%%%%%%
\section{Chiral Perturbation Theory}
\label{sec:chpt}
%%%%%%%%%%%%%%%%%%%%%%%%%%%%%%%%%%%%%%%%%%%%%%%%%%%%%%%%%%%%%%%%%%%%%%%%%%%%%%%
\noindent
The next-to-leading-order (NLO) expression in Chiral Perturbation Theory (ChPT) reads \cite{Giusti:2008vb},
\be\label{eq:chpt}
\frac{{\widetilde \Sigma}^{\rm NLO}}{\Sigma}(\Lambda_1,\Lambda_2,m) 	= 1 + \frac{m \Sigma}{(4\pi)^2 F^4}
									 \Big[ 3\, \bar l_6 + 1 - \ln(2) - 3 \ln\Big(\frac{\Sigma m}{F^2 M^2}\Big)
									+ \tilde g_\nu\left(\frac{\Lambda_1}{m},\frac{\Lambda_2}{m}\right)\Big]    
\ee
\bey
\text{with} \;\; \tilde g_\nu\left(x_1,x_2\right) 	&=& \frac{f_\nu(x_1)+f_\nu(x_2)}{2} + \frac{1}{2}\,\frac{x_1+x_2}{x_2-x_1}\,\Big[f_\nu(x_2)-f_\nu(x_1)\Big] \;, \\
f_\nu(x) 					&=& \left(x-\frac{1}{x}\right)\arctan(x)  - \frac{\pi}{2}x - \ln(x+x^3)\;,
\eey
where $F$ is the pseudo-scalar decay constant in the chiral limit, $\bar l_6$ an NLO low-energy constant (LEC) and $M$ is a scale fixed to 139.6 MeV.
This formula shows some remarkable properties. 
First of all, 
there are no chiral logs $\ln(m)$ at fixed $\Lambda$. 
Second, investigating the function $\tilde g_\nu$, one finds that ${\widetilde \Sigma}$ is a decreasing function of $\Lambda=(\Lambda_1+\Lambda_2)/2$ for any finite quark mass. 
Finally, in the chiral limit all NLO-corrections vanish in the two-flavor theory \cite{Giusti:2008vb,Smilga:1993in}. 
The dominant NNLO corrections are expected to be of the form ${\cal O}(\Lambda^2, m\Lambda, m^2)$ 
and may spoil some of the peculiar properties of the NLO expansion.
In particular, additional LECs are expected to appear and the 
 ${\cal O}(\Lambda^2)$ corrections can introduce a $\Lambda$-dependence in ${\widetilde\Sigma}$ in the chiral limit.
 
At finite lattice spacing, the chiral expansion in NLO Wilson-ChPT was carried out in the generic-small-quark-mass regime (GSM) \cite{Necco:2011vx},
\be
{\widetilde \Sigma}^{\rm NLO}_{\rm lat} = {\widetilde \Sigma}^{\rm NLO} - 32 (W_0 a)^2\frac{W_8' m}{\Lambda_1\Lambda_2} \;.
\ee
The LEC $W_8'$ was later shown to be negative \cite{Hansen:2011kk,Splittorff:2012gp}, 
which implies that also ${\widetilde \Sigma}^{\rm NLO}_{\rm lat}$ is a decreasing function of $\Lambda$ also at finite lattice spacing. 
We remark that those NLO discretization effects, and any $\Lambda$-dependence, are still absent in the chiral limit. 
The finite-volume effects have been discussed on the same footing. 
In the present study we choose the parameters such that finite-volume effects predicted by NLO ChPT are below the statistical accuracy.

%%%%%%%%%%%%%%%%%%%%%%%%%%%%%%%%%%%%%%%%%%%%%%%%%%%%%%%%%%%%%%%%%%%%%%%%%%%%%%%
\section{Simulation details}
\label{sec:simdet}
%%%%%%%%%%%%%%%%%%%%%%%%%%%%%%%%%%%%%%%%%%%%%%%%%%%%%%%%%%%%%%%%%%%%%%%%%%%%%%%
\noindent
\begin{table}
\small
\begin{center}
\setlength{\tabcolsep}{.10pc}
\begin{tabular}{@{\extracolsep{1mm}}ccccccccccc}
\hline
\hline
id  & $L/a$  &$m_\pi$&$m_\pi L$&$a$	& $R\tau_{\rm exp}$	& $R\tau_{\rm int}(m_{\pi})$ & $R\tau_{\rm int}(\nu)|_{\Lambda=85\rm MeV}$ & $\Delta_{\rm cnfg}$	&$N_{\rm cnfg}$ \\
	&	& [MeV]	&	&	[fm]		& [MDU] & [MDU]& [MDU]& [MDU] \\
\hline
 A3   & $32$& $490$  & $6.0$  & 0.0755(9)(7)	& 25	& 7	& 2(1)	& 128	& 55\\     
A4   &           & $380$  &  $4.7$  &			& 	& 5	&				& 144	& 55\\
A5   &            &$330$  & $4.0$   &			& 	& 5	& 				& 36		& 55\\
B6   & $48$   & 280    & 5.2   	&  			 & 	& 6	& 				& 24		& 50\\
\hline
E5  &$32$   &$ 440$  & $4.7$ &  0.0658(7)(7)  & 50	& 9	& 4(2)		& 96		& 92\\
F6  &$48$     &$ 310$  & $5.0$ &  			&	& 8	&				& 80		& 40\\
F7  &           &$ 270$  & $4.3$ &    			&	& 7	&				& 72		& 50 \\
G8  &$64$     &  190       &   4.1 		&   			&	& 8	&		& 48		& 22 \\
\hline
\hline
\end{tabular}
\caption{
Details of the simulation. $L$ denotes the linear size of the lattice, $a$ the lattice spacing \cite{Fritzsch:2012wq}, $m_\pi$ the pion mass, $R$ the ratio of active links in DD-HMC \cite{Luscher:2003qa} ($R=1$ in MP-HMC \cite{Marinkovic:2010eg}), 
$\tau_{\rm exp}$ and $\tau_{\rm int}$ the exponential and integrated autocorrelation time, resp., MDU molecular dynamics units, 
$\Delta_{\rm cnfg}$ the separation of configurations between subsequent measurements and $N_{\rm cnfg}$ the number of configurations on which $\nu$ is measured. 
}
\vspace{-3mm}
\label{tab:simdet}
\end{center}
\end{table}
We consider the O$(a)$-improved Wilson formulation of lattice QCD with a doublet of mass-degenerate sea quarks. 
The configurations have been generated by the CLS-initiative, the most relevant details for the present study  are depicted in Tab.~\ref{tab:simdet}, further details are found in Refs.~\cite{Fritzsch:2012wq,DelDebbio:2006cn,Marinkovic:2011pa}. 
We determine the autocorrelation time of the mode number at $\Lambda=85$ MeV for each value of the lattice spacing, which turns out to be considerably smaller than the exponential autocorrelation time related to the topology. 
Hence, with a sufficient spacing between subsequent measurements, we can safely neglect the autocorrelation. 
We also remark that $m_\pi L\ge 4$ for all ensembles, such that finite-volume effects are expected to be negligible.
The mode number is evaluated using pseudo-fermion fields $\eta_k$ and rational polynomials to approximate the spectral projector $\mathbb P_M$ to the low modes of the Dirac operator \cite{Giusti:2008vb},
\be
\nu 	= \frac{1}{N}\sum_{k=1}^N \< \left( \eta_k, \mathbb P_M \eta_k \right)\> \;, \qquad\qquad M=\sqrt{\Lambda^2+m^2}\;.  
\ee

%%%%%%%%%%%%%%%%%%%%%%%%%%%%%%%%%%%%%%%%%%%%%%%%%%%%%%%%%%%%%%%%%%%%%%%%%%%%%%%
\section{Fitting strategy and results}
\label{sec:results}
%%%%%%%%%%%%%%%%%%%%%%%%%%%%%%%%%%%%%%%%%%%%%%%%%%%%%%%%%%%%%%%%%%%%%%%%%%%%%%%
\begin{figure}[htb]
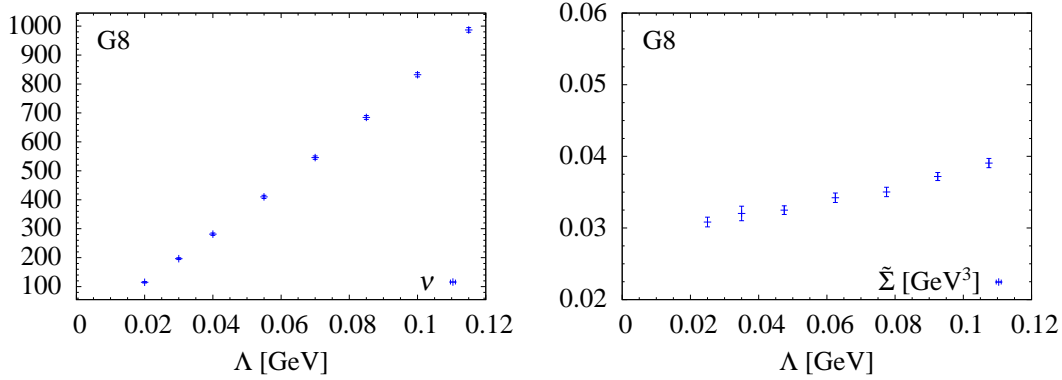

\begin{center}
\hspace{-5mm}
{\LARGE\resizebox{!}{0.35\textwidth}{\input{G8-cs2000-nu.tex}}}
\hspace{-5mm}
{\LARGE\resizebox{!}{0.35\textwidth}{\input{G8-cs2000-st.tex}}}
\vspace{-5mm}
\end{center}
\caption{Mode number $\nu$ vs.~cutoff $\Lambda$ (lhs) and effective condensate ${\widetilde \Sigma}$ vs.~$\Lambda$ (rhs) for ensemble G8 ($m_\pi\approx190$ MeV). 
All quantities are renormalized. 
Note the increasing behavior of ${\widetilde \Sigma}$ which is incompatible with NLO ChPT, and is interpreted as first indication for higher-order effects. 
Here and in other plots for a single value of $\beta$, the error shown does not include the error on the scale.
}
\label{fig:obs-vs-lam}
\end{figure}
\noindent
The range of sensible values of $\Lambda$ in the simulation is bounded from below by finite-volume and discretization effects and from above by the applicability of ChPT. 
We compute $\nu(\Lambda)$ for seven appropriate values of $\Lambda$ in the range 20 MeV $\leq\Lambda\leq$ 115 MeV.
We show results for the mode number (lhs) and for the effective condensate (rhs) for ensemble G8 in Fig.~\ref{fig:obs-vs-lam}. 
Note the statistically significant increasing dependence of ${\widetilde \Sigma}$ on $\Lambda$ which is incompatible with NLO ChPT, discussed in Sec.~\ref{sec:chpt}. 
Such a behavior is observed in all ensembles, which can be interpreted as an indication for higher-order effects. 
To fit the data in the full range of $\Lambda$ and $m$ we consider a fit form resting on NLO WChPT, but capable of accounting for higher-order effects, 
\bey\label{eq:genfit}
{\widetilde \Sigma}							&=& c_0(\Lambda) + c_1(\Lambda)m + c_2 g\left(\Lambda_1,\Lambda_2,m\right) \\
\text{with} \;\; g\left(\Lambda_1,\Lambda_2,m\right)	&=& m\left[\tilde g_\nu\left(\frac{\Lambda_1}{m},\frac{\Lambda_2}{m}\right) -3\ln\left(\frac{m}{M}\right)\right] \;.
\eey
The $\Lambda$-dependent coefficients $c_0$ and $c_1$ account for the dominant NNLO effects discussed in Sec.~\ref{sec:chpt}, in particular NLO discretization and NNLO $\Lambda^2$ and $m\Lambda$ effects. 
\begin{center}
\begin{figure}[htb]
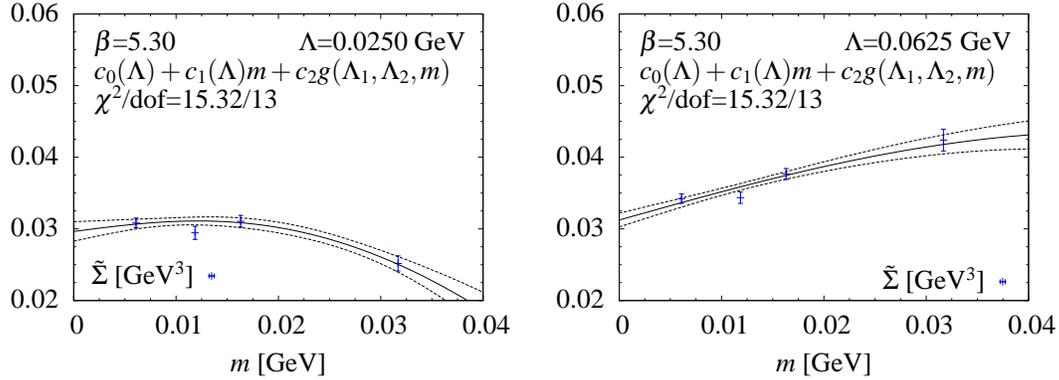

\begin{center}
\hspace{-5mm}
{\LARGE\resizebox{!}{0.35\textwidth}{\input{b5,30-g2014-lam0,0250-st.tex}}}
\hspace{-5mm}
{\LARGE\resizebox{!}{0.35\textwidth}{\input{b5,30-g2014-lam0,0625-st.tex}}}
\vspace{-1mm}
\end{center}
\caption{Effective condensate ${\widetilde\Sigma}$ vs.~the quark mass for $\Lambda=25$ MeV (lhs) and $\Lambda=62.5$ MeV (rhs) at $\beta=5.3$. 
The fit function shown follows Eq.~\eqref{eq:genfit} and has a $\chi^2$ per degree of freedom close to one. 
All fits shown in this article are performed using the double-elimination jackknife technique, accounting for the correlation of the data among different $\Lambda$. 
}
\label{fig:genfit-vs-m}
\end{figure}
\end{center}
\begin{figure}[htb]
\begin{center}
{\LARGE\resizebox{!}{0.35\textwidth}{\input{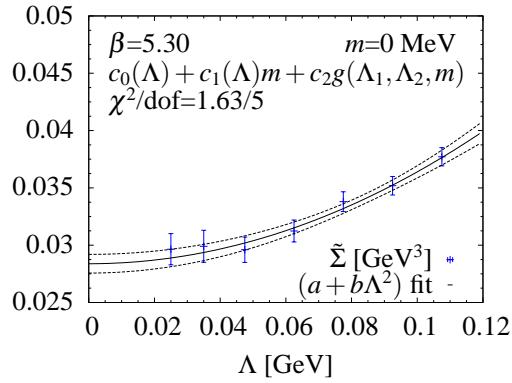}}}
\vspace{-5mm}
\end{center}
\caption{Effective condensate ${\widetilde\Sigma}$ vs.~$\Lambda$ in the chiral limit at $\beta=5.3$. 
The data points shown derive from the fit to the data following 
Eq.~\eqref{eq:genfit}. 
The plateau-like behavior for $\Lambda\lesssim50$ MeV is compatible with NLO ChPT. 
Note the significant deviation from the plateau for $\Lambda\gtrsim50$ MeV, interpreted as second indication for higher-order effects. 
The fit shown assumes an NNLO $\Lambda^2$ term in the chiral expansion of ${\widetilde\Sigma}$ and describes the data nicely. 
}
\label{fig:genfit-vs-lam}
\end{figure}
\begin{figure}[htb]
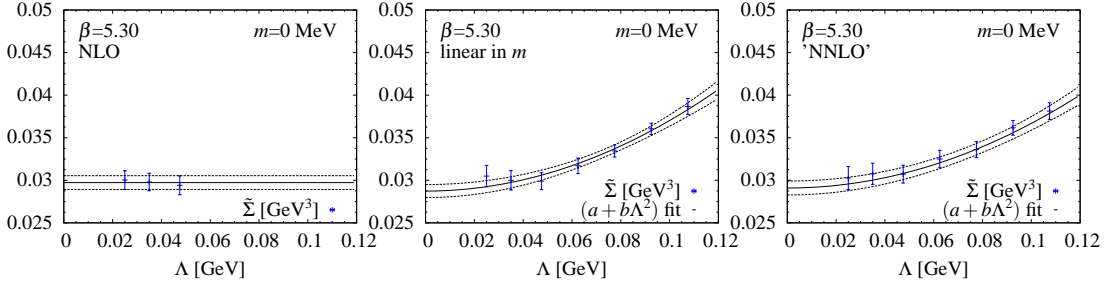

\begin{center}
\hspace{-5mm}
{\LARGE\resizebox{!}{0.26\textwidth}{\input{b5,30-m2026-c0.tex}}}
\hspace{-10mm}
{\LARGE\resizebox{!}{0.26\textwidth}{\input{b5,30-g2032-c0.tex}}}
\hspace{-10mm}
{\LARGE\resizebox{!}{0.26\textwidth}{\input{b5,30-m2212-c0.tex}}}
\vspace{-5mm}
\end{center}
\caption{Effective condensate  ${\widetilde\Sigma}$ vs.~$\Lambda$ at $\beta=5.3$ in the chiral limit for three different fitting strategies. 
Left: fit form following NLO ChPT, excluding data on heavy quark masses $m\geq20$ MeV and heavy cutoffs $\Lambda\geq50$ MeV; 
middle: fit linear in the quark mass, excluding data on heavy $m$; 
right: generic fit form (NLO plus parametrized NNLO). 
The respective fit functions are given explicitly in Tab.~\ref{tab:strats}.
All relevant $\chi^2$/d.o.f.~are close to one. 
}
\label{fig:mainfits-b5.3}
\end{figure}
\begin{figure}[htb]
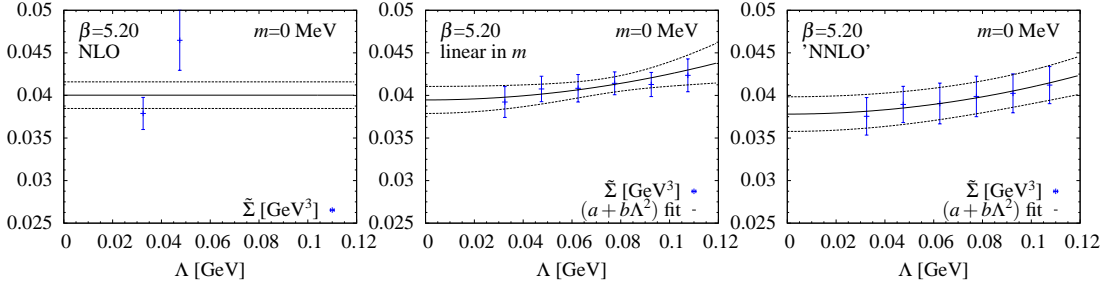

\begin{center}
\hspace{-5mm}
{\LARGE\resizebox{!}{0.26\textwidth}{\input{b5,20-m126-c0.tex}}}
\hspace{-10mm}
{\LARGE\resizebox{!}{0.26\textwidth}{\input{b5,20-g130-c0.tex}}}
\hspace{-10mm}
{\LARGE\resizebox{!}{0.26\textwidth}{\input{b5,20-n132-c0.tex}}}
\vspace{-5mm}
\end{center}
\caption{Same as Fig.~\ref{fig:mainfits-b5.3}, but for $\beta=5.2$. 
All relevant $\chi^2$/d.o.f.~are close to one. 
The poor quality of the NLO fit results from having too few data points in the relevant range; an extension of the study is currently in progress.
}
\label{fig:mainfits-b5.2}
\end{figure}
The resulting fit function for $\beta=5.3$ is shown vs.~the quark mass for two distinct values of $\Lambda$ in Fig.~\ref{fig:genfit-vs-m} and vs.~$\Lambda$ in the chiral limit in Fig.~\ref{fig:genfit-vs-lam}. 
We find that the data is described nicely and the $\chi^2$ per degree of freedom is close to one. 
The $\Lambda$-dependent coefficients $c_0$ and $c_1$ should exhibit a plateau-behavior in the range of NLO ChPT. 
Such a behavior is verified only below 50 MeV, while higher-order effects are visible above. 
This result supports the existence of a window in $\Lambda$ from which the condensate can be extracted reliably. 
Furthermore, the extracted $\Lambda$-dependencies of $c_0$ and $c_1$ are compatible with the expectations from ChPT as discussed in Sec.~\ref{sec:chpt}. 
We are thus led to specify particular functions for these coefficients, inspired by ChPT, in order to reduce the number of fit parameters. 
In particular, we consider three main fitting strategies, given in Tab.~\ref{tab:strats} together with the corresponding data ranges applied. 
\noindent
\begin{table}
\small
\caption{The three main fitting strategies used to extrapolate to the chiral limit.}
\begin{center}
\vspace{-5mm}
\begin{tabular}{cccl}
\hline
\hline
strategy name 			& $\Lambda$[MeV]	& $m$[MeV] 	& \qquad\qquad\qquad functional form 		\\
\hline
NLO 			&<~~~50 			&< 25	  	& $c_0 + c_1m + c_2g(.)
+ \;\, c_3a^2m/\Lambda^2
$\\
linear in $m$	 &< 110		 	&< 25  		& $c_0 + c_1m\qquad\quad\,
+ \;\, c_3a^2m/\Lambda^2 \quad 
+ c_4\Lambda^2
+ c_5m\Lambda
$\\
``NNLO'' 		&< 110 			&< 40  		& $\underbrace{c_0 + c_1m + c_2g(.)}_{\rm NLO}
+ \underbrace{c_3a^2m/\Lambda^2}_{\rm NLO~discret.~eff.}
+ \underbrace{c_4\Lambda^2
+ c_5m\Lambda
+ c_6m^2}_{\rm ``NNLO"}
$ \\
\hline
\hline
\end{tabular}
\label{tab:strats}
\end{center}
\vspace{-3mm}
\end{table}
The results for the three main fitting strategies are shown vs.~$\Lambda$ in the chiral limit in Figs.~\ref{fig:mainfits-b5.3} and \ref{fig:mainfits-b5.2} for $\beta=5.3$ and 5.2, respectively. 
The different strategies are well compatible towards the chiral limit, while we find discrepancies at larger quark mass as expected.
The pure NLO ChPT fit shows a systematic error of neglecting $\Lambda^2$ effects, which is almost of the order of the statistical error in the considered range of $\Lambda$. 
In addition to the three different fitting strategies, we consider several additional consistency criteria, like 
modelling/unmodelling the NNLO $\Lambda$-dependence of $c_0$ and/or $c_1$, 
using a denser set of light $\Lambda$, or
using ChPT/LQCD input for the parameter $c_2$. 
We find a good agreement between all these fitting strategies and conclude that the associated systematic errors on $\Sigma$ are reasonably well under control. 

%%%%%%%%%%%%%%%%%%%%%%%%%%%%%%%%%%%%%%%%%%%%%%%%%%%%%%%%%%%%%%%%%%%%%%%%%%%%%%%
\section{Conclusions}
\label{sec:conclusion}
%%%%%%%%%%%%%%%%%%%%%%%%%%%%%%%%%%%%%%%%%%%%%%%%%%%%%%%%%%%%%%%%%%%%%%%%%%%%%%%
\noindent
We presented preliminary results for the chiral condensate in two-flavor QCD from the Banks--Casher relation. 
Higher-order effects are observed in the investigated parameter range and appropriate fitting strategies have been discussed. 
These are found to describe the data well and point to the existence of a window in $\Lambda$ and $m$ where $\Sigma$ can be extracted reliably. 
In the chiral limit, we find $\Sigma^{1/3}=0.308(6)$ GeV and $\Sigma^{1/3}=0.336(9)$ GeV for the lattice spacings $a\approx0.066$ fm and $a\approx0.076$ fm, respectively.  
The quoted errors are estimated by quadratic summation over the statistical and systematic errors of the dimensionless condensate and the scale.
We plan to conclude the study considering a third lattice spacing and taking the continuum limit. 

%%%%%%%%%%%%%%%%%%%%%%%%%%%%%%%%%%%%%%%%%%%%%%%%%%%%%%%%%%%%%%%%%%%%%%%%%%%%%%%
\acknowledgments
%%%%%%%%%%%%%%%%%%%%%%%%%%%%%%%%%%%%%%%%%%%%%%%%%%%%%%%%%%%%%%%%%%%%%%%%%%%%%%%
\noindent
Simulations have been performed on BlueGene/Q at CINECA (CINECA-INFN agreement), Lagrange at Cilea, PAX at Desy, Zeuthen, JUROPA/JUQUEEN in J\"ulich JSC, and HLRN.
We thank these institutions for support and technical help. We are grateful to our colleagues within the CLS initiative for sharing ensembles. 
G.P.E.~and L.G.~acknowledge support by the project PRIN 2009 12-25146011-300.
S.L.~and R.S.~acknowledge support by the DFG Sonderforschungsbereich/Transregio SFB/TR9.

%\bibliographystyle{unsrt}
%\bibliography{Literature-cond-nf2}{}

\end{document}